\documentclass[showpacs,superscriptaddress,nofootinbib,amssymb,amsmath,twocolumn]{revtex4}
\usepackage{txfonts}
\usepackage[dvips]{graphicx}
\usepackage{color,latexsym,epsfig,bm,times,subfigure}
\usepackage[ps2pdf,bookmarks=true,colorlinks,linkcolor=blue,urlcolor=blue,citecolor=blue]{hyperref}
\renewcommand\a{\alpha}
\renewcommand\b{\beta}

\newcommand\g{\gamma}
\newcommand\z{\zeta}

\newcommand\p{\pi}

\newcommand\f{\phi}

\newcommand\D{\Delta}

\newcommand\J{\Psi}

\newcommand{\eq}[1]{Eq.~(\ref{#1})}

\newcommand{\lan}{\langle}
\newcommand{\ran}{\rangle}

\newcommand{\rp}{{\rm RP}}

\renewcommand{\part}{{\rm part}}

\begin{document}

\title{Possible observables for the chiral electric separation effect in Cu + Au collisions}

\author{Guo-Liang Ma}
\email[]{glma@sinap.ac.cn}
\affiliation{Shanghai Institute of Applied Physics, Chinese Academy of Sciences, Shanghai 201800, China}

\author{Xu-Guang Huang}
\email[]{huangxuguang@fudan.edu.cn}
\affiliation{Physics Department and Center for Particle Physics and Field Theory, Fudan University, Shanghai 200433, China}


\begin{abstract}
The quark-gluon plasma (QGP) generated in relativistic heavy-ion collisions could be locally parity-odd. In parity-odd QGP, the electric field may induce a chiral current which is called the chiral electric separation effect (CESE). We propose two possible observables for CESE in Cu + Au collisions: The first one is the correlation $\zeta_{\alpha\beta}=\langle \cos[2(\phi_\alpha+\phi_\beta-2\Psi_{\rm RP})]\rangle$; the second one is the charge-dependent event-plane angle $\Psi^{q}_2$ with $q=\pm$ being charge. Nonzero $\Delta\zeta=\zeta_{opp}-\zeta_{same}$ and $\Delta\Psi=\langle|\Psi_2^+-\Psi_2^-|\rangle$ may signal the CESE in Cu + Au collisions. Within a multiphase transport model, we study how the final state interaction affects these observables. We find that the correlation $\gamma_{\alpha\beta}=\langle\cos(\phi_{\alpha}+\phi_{\beta}-\Psi_{\rm RP})\rangle$ is sensitive to the out-of-plane charge separation caused by the chiral magnetic effect and to the in-plane charge separation caused by the in-plane electric field, but it is not sensitive to the CESE. On the other hand, $\Delta\zeta$ and $\Delta\Psi$ are sensitive to the CESE. Therefore, we suggest that the future experiments measure the above observables in Cu+Au collisions in order to disentangle different chiral and charge separation mechanisms.
\end{abstract}


\maketitle

\section{Introduction}
\label{sec:intro}

Relativistic heavy-ion collisions generate not only extremely hot quark-gluon plasma (QGP) but also extremely large magnetic fields due to the fast motion of the colliding ions. Recent detailed calculations revealed that the maximum magnetic fields in Au + Au collisions at energies currently available at the BNL Relativistic Heavy Ion Collider (RHIC) can reach $5 m_\p^2 \sim 10^{18}$ Gauss while in Pb + Pb collisions at energies available at the CERN Large Hadron Collider (LHC) they can reach $60 m_\p^2 \sim 10^{19}$ Gauss~\cite{Skokov:2009qp,Voronyuk:2011jd,Deng:2012pc,Bzdak:2011yy,Bloczynski:2012en}. Under such large magnetic fields, some novel quantum phenomena can possibly happen. The most intriguing ones are the so-called chiral magnetic effect (CME)~\cite{Vilenkin:1980fu,Kharzeev:2004ey,Kharzeev:2007tn,Kharzeev:2007jp,Fukushima:2008xe} and chiral separation effect (CSE)~\cite{Son:2004tq,Metlitski:2005pr}: They occur in parity- and charge-odd regions in QGP and can result in charge and chirality separations along the direction of the magnetic field, respectively. Recent experimental measurements of the charge azimuthal correlation~\cite{Voloshin:2004vk},
\begin{eqnarray}
\label{gam}
\g_{\a\b}=\lan\cos(\f_\a+\f_\b-2\J_\rp)\ran,
\end{eqnarray}
where $\phi_{\alpha}$ and $\phi_{\beta}$ are the emission azimuthal angles of particles with charges $\alpha$ and $\beta$ and $\J_\rp$ is the reaction plane angle,
showed some consistent features with the expectation of CME at both RHIC and LHC energies~\cite{Abelev:2009ac,Abelev:2012pa,Adamczyk:2014mzf}. Another important experimental test of CME and CSE is the observation of the charge asymmetry in the elliptic flow of pions~\cite{Wang:2012qs,Ke:2012qb} that is consistent with the expectation of a chiral magnetic wave (CMW)~\cite{Kharzeev:2010gd,Burnier:2011bf} --- a collective mode arising due to the interplay between CME and CSE in the presence of the magnetic field. However, one must notice that there are other backgrounds that contribute to the experimental observables, see Refs.~\cite{Wang:2009kd,Bzdak:2009fc,Bzdak:2010fd,Liao:2010nv,Pratt:2010zn,Schlichting:2010qia,Bzdak:2012ia,Bloczynski:2013mca,Liao:2014ava} for discussions.

Heavy-ion collisions can also generate strong electric fields due to the event-by-event fluctuation of the proton positions in the ions~\cite{Deng:2012pc,Bzdak:2011yy} or due to asymmetric colliding geometry (for example, in Cu + Au collisions~\cite{Deng:2014uja,Hirono:2012rt,Voronyuk:2014rna}). It was proposed that the electric fields may also induce chiral current and chiral separation in QGP, which is called the chiral electric separation effect (CESE)~\cite{Huang:2013iia,Jiang:2014ura,Pu:2014cwa,Pu:2014fva}. It was also proposed that the Cu+Au collisions may provide us a good chance to detect CESE because there is a strong electric field directing from Au to Cu due to the charge asymmetry between Au and Cu nuclei~\cite{Huang:2013iia}. Recently, the electromagnetic fields in Cu + Au collisions were studied in details, and it was found that the large in-plane charge separation effect (in-plane CSE) induced by the strong in-plane electric fields could strongly suppress or even reverse signs of $\g_{\a\b}$~\cite{Deng:2014uja}. 

In this paper, we propose two observables that are designed solely for the detection of CESE. In addition, it is crucial to take the final state interactions into account for any model calculations in order to link the initial anomalous transports to the experimental data since heavy-ion collisions undergo complicated dynamical evolutions which involve many final interactions. We study the effects of the final state interactions by using a multi-phase transport (AMPT) model which successfully describes the main evolution stages of heavy-ion collisions. By introducing appropriate initial dipolar or quadrupole charge distributions, the AMPT model can successfully describe both the charge azimuthal correlation $\g_{\a\b}$~\cite{Ma:2011uma,Shou:2014zsa} and the charge asymmetry of the pion elliptic flow~\cite{Ma:2014iva} in Au + Au collisions at the top RHIC energy. In this work, we introduce different kinds of initial charge separations which are used to mimic different initial chiral effects into the initial condition of the AMPT model, and predict some observables which can be used to test whether these effects can be observed in the final state of Cu+Au collisions at $\sqrt{s_{_{\rm NN}}}$ = 200 GeV.

The paper is organized as follows. In Sec.~\ref{sec:model} we setup the numerical simulations, in Sec.~\ref{sec:results} we show our main results. Finally we summarize and discuss in Sec.~\ref{sec:summary}. Throughout this paper, we use the natural units $\hbar=k_B=c=1$.

\section{General Setup}
\label{sec:model}
\subsection{AMPT model}

The AMPT model with a string melting scenario is utilized in this work~\cite{Lin:2004en}. The AMPT model is a dynamical transport model which consists of four main components: initial condition, parton cascade, hadronization, and hadronic rescatterings. The initial condition, which includes the spatial and momentum distribution of participant matter, minijet partons production and soft string excitations, is obtained through the HIJING model~\cite{Wang:1991hta,Gyulassy:1994ew}. The parton cascade starts the partonic evolution with a quark-anti-quark plasma from the melting of strings. Parton scatterings are modelled by Zhang's parton cascade (ZPC), which currently only includes two-body elastic parton scatterings using cross sections from pQCD with screening masses~\cite{Zhang:1997ej}. The parton interaction cross section is set as 10 mb in our work, by which the model has shown good abilities to describe many key experimental observables at RHIC~\cite{Chen:2004dv, Zhang:2005ni, Chen:2006ub,Ma:2010dv,Lin:2002gc,Ko:2013mf}. A quark coalescence model is then used to combine partons into hadrons when the system freezes out. The evolution dynamics of the hadronic matter is described by a relativistic transport (ART) model~\cite{Li:1995pra}. Because the current implementation of the ART model does not conserve the electric charge, we in this study consider only resonance decays and hadronic scatterings are switched off to ensure the charge conservation.

\subsection{Introducing the initial charge separations into the AMPT model}
\label{sec:jetrec}

Let us consider the chiral magnetic effect (CME) in Cu + Cu collisions first. This can serve as a test of the AMPT model. The coordinate system is set up so that the $x$-axis is in the reaction plane, i.e., $\Psi_{\rm RP}=0$, and the $y$-axis is perpendicular to the reaction plane with the $z$-axis being the direction of the projectile nucleus. To mimic the CME, we introduce an initial charge separation into the initial state of the AMPT model, since the charges are not separated but distributed randomly in the normal AMPT model. To separate a percentage of the initial charges, we follow the procedure of a global charge separation scenario, which has been employed in Ma and Zhang's previous work~\cite{Ma:2011uma}.  We switch a percentage of the downward moving $u$ quarks with those of the upward moving $\bar{u}$ quarks in such a way that the total momentum is conserved, and likewise for $\bar{d}$ and $d$ quarks, where the percentage should presumably depend on the impact parameter $b$ or centrality, because the magnitude of the averaged magnetic field is $b$ dependent.
\begin{figure}
\includegraphics[scale=0.45]{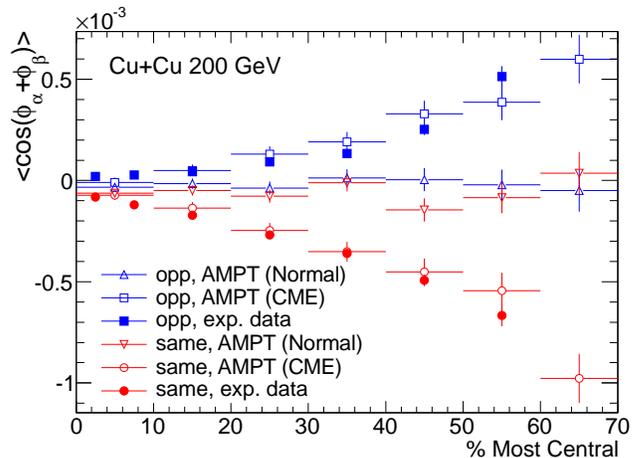}
\caption{(Color online) Centrality dependence of $\langle\cos(\phi_{\alpha}+\phi_{\beta})\rangle$ in Cu+Cu collisions at $\sqrt{s_{_{\rm NN}}}$=200 GeV. The open symbols represent the results from the normal AMPT model without anomalous effects and the AMPT model with an initial CME-like charge separation, and the solid symbols represent the experimental data~\cite{Abelev:2009ac}.}
 \label{fig-CuCu}
\end{figure}

In Fig.~\ref{fig-CuCu} we show our AMPT results of the charge azimuthal correlation $\g_{\a\b}$ (from the normal AMPT model and the AMPT model with the initial CME-like charge separation) as well as the experimental data for Cu+Cu collisions at $\sqrt{s_{_{\rm NN}}}$ = 200 GeV. We find that, to fit the Cu+Cu experimental data, the percentage of initial CME-like charge separation $f\%$ should be proportional to the impact parameter $b$ with a slope of 1.56, i.e. $f=1.56b/{\rm fm}$. This is consistent with the fact that the averaged magnetic field is proportional to $b$~\cite{Deng:2012pc,Bzdak:2011yy}. Comparing with the normal AMPT result, the AMPT result with the initial CME-like charge separation can well describe Cu+Cu data measured by the solenoidal tracker at RHIC (STAR) experiment. 
\begin{figure*}
\includegraphics[scale=0.8]{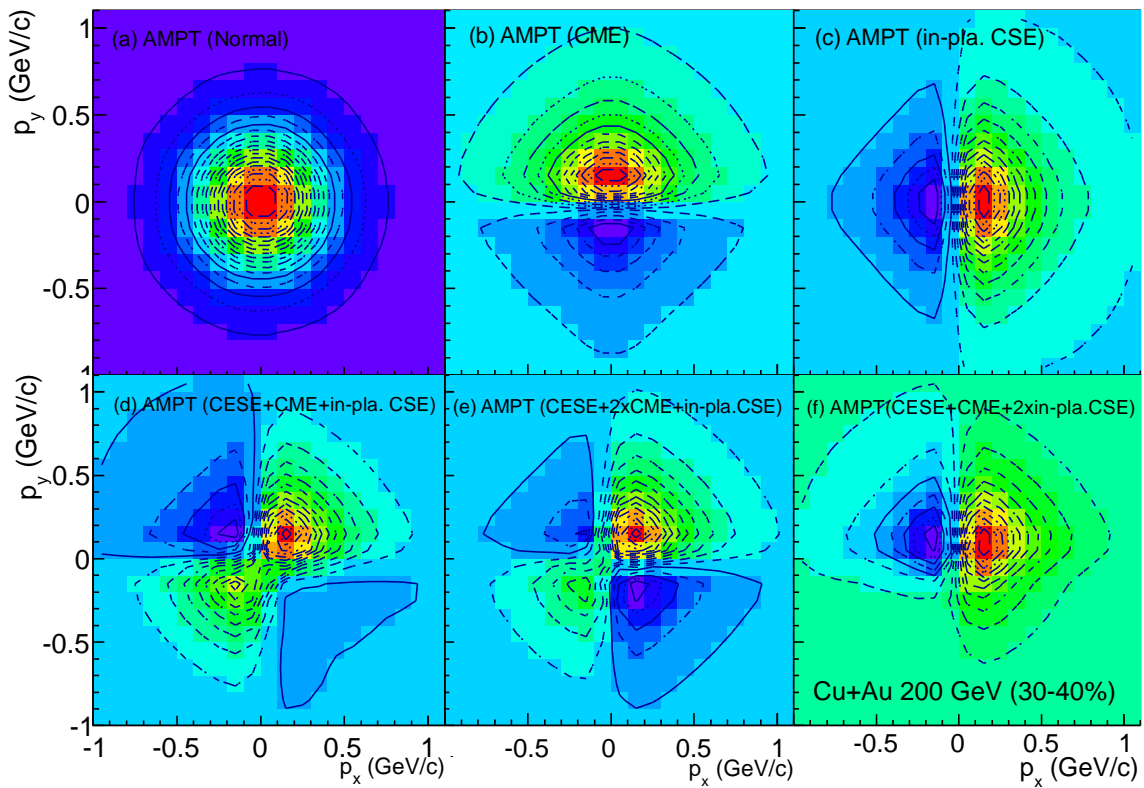}
\caption{(Color online) The net electric charge distributions in the transverse momentum plane in the initial partonic states for different settings of AMPT models for centrality bin of 30-40\% in Cu+Au collisions $\sqrt{s_{_{\rm NN}}}$=200 GeV. (a) Normal setting without any anomalous effect. (b) AMPT with out-of-plane charge separation which mimic the chiral magnetic effect (CME). (c) AMPT with in-plane charge separation effect (in-plane CSE) caused by the in-plane electric field. (d) AMPT with in-plane CSE plus a quadrupolar distribution due to chiral electric separation effect (CESE) and CME. (e) The same as (d) but with the strength of CME doubled. (f) The same as (d) but with the strength of the in-plane CSE doubled.}
 \label{fig-changedis}
\end{figure*}

\begin{figure}
\includegraphics[width=8.5cm]{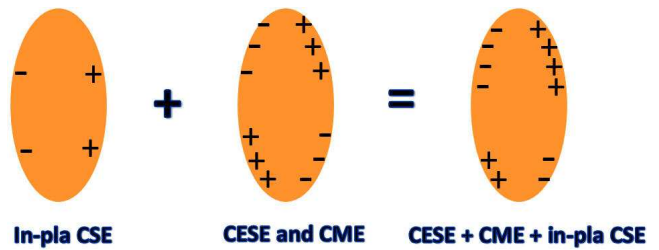}
\caption{(Color online) Illustration of the charge configuration of chiral magnetic effect (CME) plus chiral electric separation effect (CESE) plus in-plane charge separation effect (in-plane CSE) in Cu + Au collisions.}
 \label{illu}
\end{figure}

We now turn to study the Cu + Au systems. The coordinate system of Cu + Au is set up similarly to that of Cu+Cu, but the Au nucleus is set to the left of the Cu nucleus, i.e., the direction of the total electric field is rightward. Figures~\ref{fig-changedis} (a)-(f) show the net electric charge distributions in the transverse momentum space of the initial partonic states in the AMPT models with different kinds of initial charge separations for centrality bin of 30-40\% in Cu+Au collisions $\sqrt{s_{_{\rm NN}}}$=200 GeV.  After introducing the CME effect into the normal AMPT model, the net electric charge distribution is changed from that in Fig.~\ref{fig-changedis} (a) to that in Fig.~\ref{fig-changedis} (b). Due to the strong electric fields in the in-plane direction in Cu + Au collisions, there may appear also the in-plane charge separation effect (in-plane CSE)~\cite{Deng:2014uja}. To simulate the in-plane CSE, we take an analogous setup as above by switching the $p_{x}$ values of a percentage of the leftward moving $u$ quarks with those of the rightward moving $\bar{u}$ quarks, and likewise for $\bar{d}$ and $d$ quarks. The $b$ dependence of the percentage is assumed to be the same as that for the out-of-plane separation. The corresponding net electric charge distribution is shown in Fig.~\ref{fig-changedis} (c). Now we take the chiral electric separation effect (CESE) into account. We consider the situation where the CESE, CME and in-plane CSE happen simultaneously (denoted as CESE+CME+in-plane CSE)~\cite{Huang:2013iia}: For the quarks with $p_y>0$ we switch a percentage $f$\% of the leftward moving $u$ quarks with those of the rightward moving $\bar{u}$ quarks (likewise for $\bar{d}$ and $d$ quarks); while for quarks with $p_y<0$ we switch half of the percentage (i.e. $0.5 f$\%) for the leftward moving $\bar{u}$ ($d$) quarks with those of the rightward moving $u$ ($\bar{d}$) quarks. For the initial charge separation $f\%$, we apply the same $b$-dependent initial charge separation as that of Cu+Cu collisions at $\sqrt{s_{_{\rm NN}}}$=200 GeV. In this case, this $b$-dependent percentage is expected to provide a lower limit for the initial charge separation percentage, since the Cu+Au system has a little bit stronger magnetic field than the Cu+Cu system at the same impact parameter~\cite{Deng:2014uja}. The net effect is equivalent to a configuration with a in-plane dipole plus a quadrupole as illustrated in Fig.~\ref{illu}. The net electric charge distribution for CESE+CME+in-plane CSE is shown in Fig.~\ref{fig-changedis} (d). To study the case where the CME is much stronger, we do the CME switching once more after the above CESE+CME+in-plane CSE switching, which is shown in Fig.~\ref{fig-changedis} (e). If the in-plane CSE effect is much larger, we do the in-plane CSE switching once more after the CESE+CME+in-plane CSE switching, see Fig.~\ref{fig-changedis} (f). In these ways, we have six different kinds of initial charge distributions which mimic different initial chiral effects. We use them as the different inputs for the initial condition of the AMPT model, and then we extract various hadronic observables to test whether the initially embedded effects can survive after the final state interactions. The results are presented in the following.

\section{Results and Discussions}
\label{sec:results}
\subsection{Correlation $\gamma_{\a\b}$}
\label{subsec:gamma}

\begin{figure*}
\includegraphics[scale=0.8]{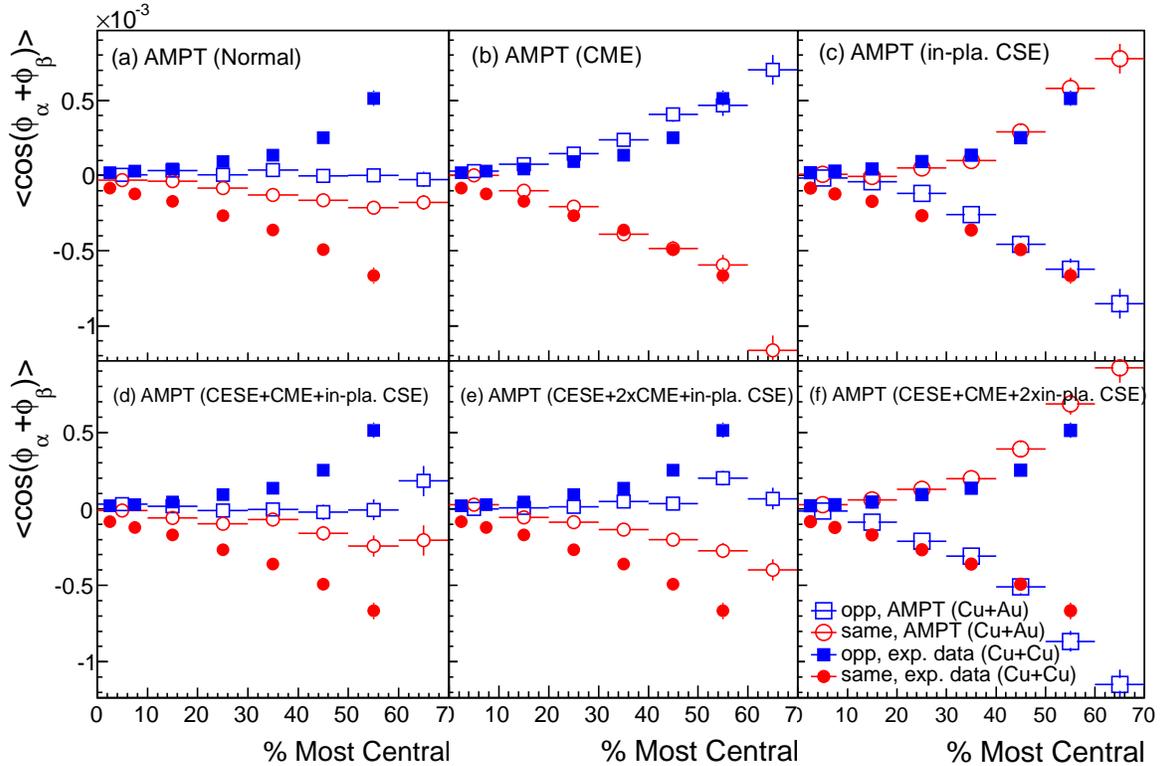}
\caption{(Color online) Centrality dependence of  $\g_{\a\b}=\langle\cos(\phi_{\alpha}+\phi_{\beta})\rangle$ from different initial settings of AMPT models (open symbols) in Cu+Au collisions at $\sqrt{s_{_{\rm NN}}}$=200 GeV. Also shown are the experimental data for Cu+Cu collisions at $\sqrt{s_{_{\rm NN}}}$=200 GeV (solid symbols)~\cite{Abelev:2009ac}. Different panels are in one-to-one correspondence with Fig.~\ref{fig-changedis}.
}
\label{fig-cos}
\end{figure*}
We first consider the correlation $\g_{\a\b}$ as defined in \eq{gam}.
The centrality dependence of $\g_{\a\b}$ in Cu+Au collisions at $\sqrt{s_{_{\rm NN}}}$=200 GeV from different settings of the AMPT model are shown in Figs.~\ref{fig-cos} (a)-(f), where the centrality bins are defined by using different ranges of impact parameter. Also shown is the experimental data from Cu + Cu collisions which are used as the baseline for comparison. Figure~\ref{fig-cos} (a) shows that the normal AMPT model without any initial charge separation gives a zero opposite-charge correlation and a negative same-charge correlation which have much smaller magnitudes than the experimental data for Cu+Cu collisions at $\sqrt{s_{_{\rm NN}}}$=200 GeV. After introducing the chiral magnetic effect (CME) into the AMPT model, as shown in Fig.~\ref{fig-cos} (b), the opposite-charge correlation becomes positive and the same-charge correlation becomes more negative, which present similar magnitudes as the Cu+Cu data. It is consistent with the previous AMPT works about Au+Au collisions~\cite{Ma:2011uma} and the above results about Cu+Cu collisions. If the initial charges are separated along the reaction plane direction [mimicking the in-plane charge separation effect (in-plane CSE) caused by the strong in-plane electric field], comparing with the CME case, opposite-charge and same-charge correlations reverse their signs, i.e. a negative opposite-charge correlation and a positive same-charge correlation are observed and their magnitudes are comparable to the Cu+Cu data, as shown in Fig.~\ref{fig-cos} (c). This result is consistent with recent study in Ref.~\cite{Deng:2014uja} and more discussions can be found there. It is also consistent with the charge asymmetry of direct flow $v_{1}$ suggested in Ref.~\cite{Hirono:2012rt}. If the chiral electric separation effect (CESE), CME, and in-plane CSE happen together in the initial stage [Fig.~\ref{fig-cos} (d)], both the opposite-charge and same-charge correlations are strongly suppressed to the levels close to the normal AMPT model [Fig.~\ref{fig-cos} (a)], which indicates the three effects almost cancel out. This can be understood by considering a limiting case in which the emitting angles for positive charges are $\p/4$ and $5\p/4$ while for negative charges they are $3\p/4$ and $7\p/4$. In this special limit, one can easily check that both $\g_{same}$ and $\g_{opp}$ vanish. If the initial strength of the CME-like charge separation gets larger, e.g. if it is doubled (denoted as CESE+2$\times$CME+in-plane CSE), the magnitudes for both opposite-charge and same-charge correlations are between those for the CME and for CESE+CME+in-plane CSE, which is shown in Fig.~\ref{fig-cos} (e). On the other hand, if we double the strength of the initial in-plane charge separation, e.g. in the case of CESE+CME+2$\times$in-plane CSE [Fig.~\ref{fig-cos} (f)], the in-plane CSE will dominate the final signal, which shows similar magnitudes as those for the in-plane CSE effect only [Fig.~\ref{fig-cos} (c)]. From these simulations, we find that the correlation $\g_{\a\b}$ is very sensitive to CME and in-plane CSE but not sensitive to CESE. To test the chiral electric separation effect, we need to look for other observables.

\subsection{Two observables for CESE}
\label{subsec:cese}

\begin{figure}
\includegraphics[scale=0.45]{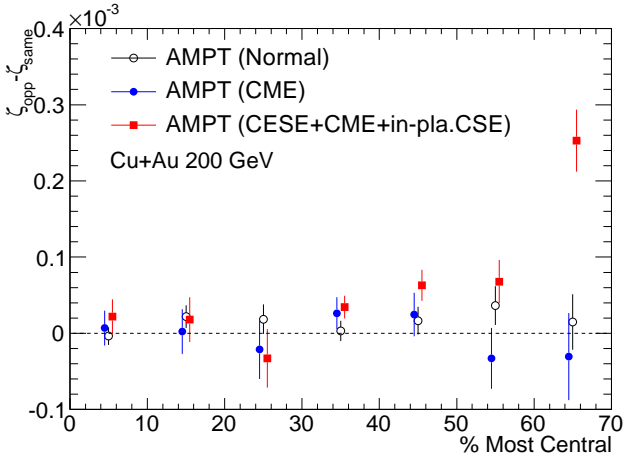}
\caption{(Color online) Centrality dependence of  the difference between opposite-charge ($\zeta_{opp}$) and same-charge ($\zeta_{same}$) correlations of $\zeta_{\alpha\beta}=\langle\cos[2(\phi_{\alpha}+\phi_{\beta})]\rangle$ for three different initial settings of AMPT models in Cu+Au collisions at $\sqrt{s_{_{\rm NN}}}$=200 GeV. Some points are slightly shifted along the horizonal axis for clearer representation.
}
 \label{fig-cos2}
\end{figure}
In this subsection, we propose two observables for the chiral electric separation effect (CESE) in Cu + Au collisions. Let us first consider the following charge azimuthal correlation:
\begin{eqnarray}
\label{zet}
\zeta_{\alpha\beta}=\langle\cos[2(\phi_{\alpha}+\phi_{\beta}-2\Psi_{\rm RP})]\rangle.
\end{eqnarray}
We will set $\J_{\rm RP}=0$ in the following. For a CESE + CME induced charge distribution [see Fig.~\ref{illu} for illustration], suppose the positive charges fly out along the azimuthal angles $\psi_{c}$ and $\psi_{c}+\pi$ and the negative charges fly out along $-\psi_{c}$ and $-\psi_{c}+\pi$, where $\psi_{c}$ is a deformation angle due to the CESE effect. Thus $\zeta_{same}\sim \langle\cos(4\psi_{c})\rangle$ should be in general different from  $\zeta_{opp}\sim 1$. On the other hand, for purely dipolar charge distribution (due to the in-plane CSE or due to the CME), $\zeta_{same}=\zeta_{opp}\sim \langle\cos(4\psi_{d})\rangle$, where $\psi_{d}$ is the dipolar angle ($\psi_d=0$ for in-plane dipole and $\psi_d=\p/2$ for out-of-plane dipole). Therefore, $\Delta\zeta$, the difference between $\zeta_{opp}$ and $\zeta_{same}$, can be an observable for the CESE since the in-plane CSE or the CME does not contribute to it. In real experiments, the dominant background of $\z_{\a\b}$ may be the elliptic flow $v_2$ because, if we turn off all the anomalous effects, $\z_{\a\b}\sim v_2^2$. Another background may be the transverse momentum conservation~\cite{Pratt:2010zn} which causes a contribution $\z_{\a\b}\propto v_4/M$ with $v_4$ the fourth harmonic flow and $M$ the multiplicity. In $\D\z$ these charge-blind $v_2$ and $v_4$ backgrounds are subtracted. However, there may remain other backgrounds, for example, due to the local charge conservation~\cite{Schlichting:2010qia} (which leads to $\z_{opp}\propto v_4/M$ and $\z_{same}\sim 0$ and thus $\D\z\propto v_4/M$) or due to the chiral magnetic wave induced quadrupole. Because we do not encode these effects into our AMPT model, the following result is supposed not to be related to the local charge conservation and chiral magnetic wave. In Fig.~\ref{fig-cos2} we present the centrality dependence of $\D\zeta$ in Cu + Au collisions from three different initial settings in the AMPT model. One can see that the normal AMPT case and the CME can not yield a visible signal for all centrality bins, while the result for CESE+CME+in-plane CESE case shows an increasing trend of $\D\zeta$ from central to peripheral Cu+Au collisions, where the signal becomes very clear in the most peripheral centrality bin.

\begin{figure}
\includegraphics[scale=0.45]{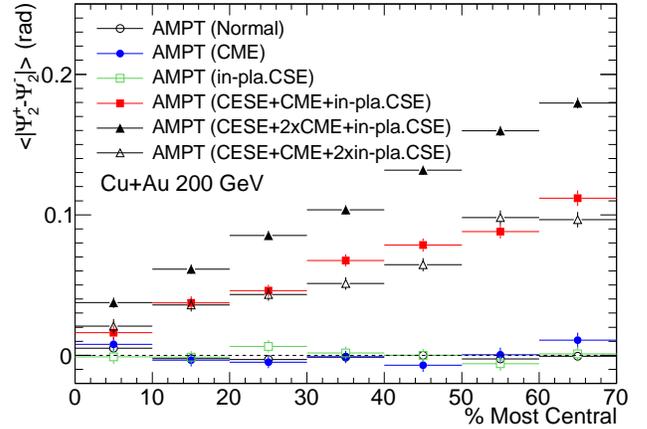}
\caption{(Color online) Centrality dependence of the difference between event plane angles reconstructed from the positively charged hadrons and negatively charged hadrons, $\langle|\Psi_{2}^{+}-\Psi_{2}^{-}|\rangle$, for six initial settings of AMPT models in Cu+Au collisions at $\sqrt{s_{_{\rm NN}}}$=200 GeV.
}
 \label{fig-psidiff}
\end{figure}
As discussed above, the chiral electric separation effect (CESE) combined with the chiral magnetic effect (CME) can lead to a quadrupolar charge distributions in Cu + Au collisions, see Fig.~\ref{illu}. Therefore, we propose another observable for CESE: $\D\Psi=\langle|\Psi_{2}^{+}-\Psi_{2}^{-}|\rangle$, where $\Psi_{2}^{+}$ and $\Psi_{2}^{-}$ are the second harmonic angles of the event planes reconstructed by final positively charged hadrons and negatively charged hadrons. If we follow the above assumption for the CESE, then $\D\J\sim 2\langle\psi_{c}\rangle$. In Fig.~\ref{fig-psidiff} we show the centrality dependence of $\D\Psi$ for six different initial settings of the AMPT model in Cu+Au collisions. As we expected, once the CESE happens, it leads to a sizable $\D\Psi$ which shows a linear dependence on the centrality. More importantly, for those initial settings which do not include the CESE, we do not observe a nonzero $\D\J$ for the whole centrality window. It should be mentioned that our results are based on the fact that in each event we know which side is the Au nucleus and which side is the Cu nucleus in the transverse plane, i.e. the direction of the electric field. In real experiments, it becomes very complex to identify the relative locations of Au and Cu and one has to reconstruct the first harmonic event plane in order to distinguish the Au side from the Cu side, which will be hopefully achieved by using some event plane detectors such as zero degree calorimeters (ZDCs) in future experiments.

\section{Summary}
\label{sec:summary}

In summary, we have introduced various initial charge separation effects, including the chiral magnetic effect (CME), in-plane charge separation effect (in-plane CSE), and chiral electric separation effect (CESE), into the AMPT model to study how the final state interactions render the initial charge separation effects.To distinguish these initial effects, three possible observables are tested in Cu+Au collisions at $\sqrt{s_{_{\rm NN}}}$=200 GeV. The charge azimuthal correlation
$\langle\cos(\phi_{\alpha}+\phi_{\beta}-2\Psi_{\rm RP})\rangle$ is sensitive to the CME and the in-plane CSE but not to the CESE. The CME results in a positive opposite-charge correlation and a negative same-charge correlation, while the in-plane CSE reverse their signs relative to those for the CME case.  However, the $\zeta_{opp}-\zeta_{same}$ is an observable that is sensitive to the CESE but not sensitive to CME and in-plane CSE which makes it be a possible observable for the detection of CESE. The difference between the second harmonic angles of event planes reconstructed by final positive-charge hadrons or negative-charge hadrons, $\langle|\Psi_{2}^{+}-\Psi_{2}^{-}|\rangle$, is also sensitive to the CESE and is sizable in very peripheral Cu+Au collisions. We thus propose that future experiments can take advantage of Cu + Au collisions to measure these observables in order to identify or distinguish different chiral and charge separation mechanisms.

\section*{ACKNOWLEDGMENTS}

Discussions with J. Liao and G. Wang are appreciated. G.-L. M. is supported by the Major State Basic Research Development Program in China under Grant No. 2014CB845404, the National Natural Science Foundation of China under Grants No. 11375251, No. 11175232, and No. 11421505, and the Knowledge Innovation Program of Chinese Academy of Sciences under Grant No. KJCX2-EW-N01. X. -G. H. is supported by Shanghai Natural Science Foundation under Grant No. 14ZR1403000 and Fudan University under Grant No. EZH1512519. We also acknowledge the support from the Innovation Fund of Key Laboratory of Quark and Lepton Physics (MOE), CCNU (Grant No. QLPL2011P01 and QLPL20122).

\end{document}